\documentclass[aps,pra,letterpaper,10pt,twocolumn,superscriptaddress]{revtex4-2}
\usepackage[utf8]{inputenc}
\usepackage[T1]{fontenc}
\usepackage[english]{babel}
\usepackage{amsmath} 
\usepackage{amssymb} 
\usepackage{amsfonts} 
\usepackage{graphicx} 
\usepackage{mathrsfs} 
\usepackage{bm} 
\usepackage{color} 
\usepackage{hyperref} 
\hypersetup{colorlinks=true,linkcolor=blue,citecolor=cyan} 
\usepackage{physics} 
\usepackage{braket} 
\usepackage{bbold} 
\usepackage{siunitx} 
\usepackage{xcolor} 
\usepackage{times} 
\usepackage{caption}
\usepackage{ragged2e}

\usepackage{float}
\usepackage{physics}
\DeclareUnicodeCharacter{0301}{'}

\def\ee{\operatorname{e}}

\begin{document}

\title{Modeling Nonlinear Exciton-Polariton Microcavities Using the Transfer Matrix Method: Spontaneous Symmetry Breaking and Self-Focusing}

\author{Karol Sajnok}
\email{ksajnok@cft.edu.pl}
\affiliation{Center for Theoretical Physics, Polish Academy of Sciences, Aleja Lotników 32/46, 02-668 Warsaw, Poland}
\affiliation{Institute of Physics, Polish Academy of Sciences, Aleja Lotników 32/46, 02-668 Warsaw, Poland}
\affiliation{Institute of Theoretical Physics, University of Warsaw, ul. Pasteura 5, 02-093 Warsaw, Poland}

\author{Michał Matuszewski}
\email{mmatuszewski@cft.edu.pl}
\affiliation{Center for Theoretical Physics, Polish Academy of Sciences, Aleja Lotników 32/46, 02-668 Warsaw, Poland}
\affiliation{Institute of Physics, Polish Academy of Sciences, Aleja Lotników 32/46, 02-668 Warsaw, Poland}

\date{\today}
\begin{abstract}
\fontsize{9pt}{11pt}\selectfont 
 The Transfer Matrix Method (TMM) is a widely used technique for modeling linear propagation of electromagnetic waves through stratified layered media. However, since its extension to inhomogeneous and nonlinear systems is not straightforward, much more computationally demanding methods such as Finite-difference time-domain (FDTD) or Method of lines (MoL) are typically used. In this work, we extend the TMM framework to incorporate the effects of nonlinearity. We consider the case where strong coupling between excitons (electron-hole pairs) and photons leads to the formation of exciton-polaritons. This extension is crucial for accurately simulating the behavior of light in polariton microcavities, where  nonlinearities arising from exciton-exciton interactions play a key role. We perform efficient simulations of light transmission and reflection in a multidimensional system using the plane wave basis. Additionally, we compare our extended TMM approach with the state-of-the-art admittance transfer method, and highlight the computational advantage of extended TMM for large-scale systems. The extended TMM not only provides a robust and computationally efficient numerical framework, but also paves the way for the development of future low-power nonlinear optical devices, polariton-based photonic circuits, and quantum photonic technologies.
\end{abstract}

\maketitle
\footnotesize
 
\section{Introduction}
\fontsize{9pt}{11pt}\selectfont 
The study of exciton-polaritons in semiconductor microcavities has garnered significant attention since the end of the 20th century \cite{weisbuch1992observation,kavokin2003cavity, shelykh2009polariton, zayats2005nano, carusotto2013quantum, low2017polaritons} due to their hybrid nature, which combines both light and matter properties. Exciton-polaritons \cite{hopfield1958theory} arise from the strong coupling between excitons (bound electron-hole pairs) and photons confined within a microcavity. Due to their inherent interactions, polaritons hold great promise for realizing nonlinear optical effects at very low power thresholds, which is crucial for the development of future photonic and quantum computing devices \cite{kavokin2022polariton, ghosh2020quantum, matuszewski2021energy, opala2022training, basov2016polaritons,opala2023harnessing,ballarini2020polaritonic,mirek2021neuromorphic}.

One of the most important aspects of studying polaritonic systems is solving Maxwell's equations in the context of complex, nonlinear optical media such as microcavities. These equations govern the propagation of electromagnetic fields in materials with varying properties, including the spatially dependent refractive index, and nonlinear blueshift resulting from polariton-polariton interactions \cite{kavokin2017microcavities}. Various methods have been developed to solve Maxwell's equations in structured media, ranging from analytical approaches to numerical simulations. Among these methods, the Transfer Matrix Method (TMM) has proven to be a versatile and efficient tool for calculating the transmission and reflection of light through linear stratified media \cite{kavokin2017microcavities}.

In its original form, the TMM is well-suited for solving linear, stratified problems, where the refractive index of each infinitely extending layer is constant and independent of the electromagnetic field intensity \cite{born2013principles}. To address either the spatially-varying effective refractive index, or the nonlinear response of the medium, which also leads to spatial inhomogeneity in the case of inhomogeneous excitation, modifications to the standard TMM are necessary. In the case of quantum-well polaritons, the nonlinear layer can be usually approximated by a thin exciton quantum well with thickness much smaller than the wavelength of light. Still, transverse variance of the effective nonlinear refractive index change leads to convolutional relations, and the transfer matrix method must be extended to incorporate mixing of Fourier modes with different momenta. This extension accounts for polarization effects, such as the TE-TM splitting, which must be incorporated to accurately capture the optical response of these systems \cite{kavokin2007polariton}. Furthermore, when considering polaritonic systems, strong coupling of photons to excitons complicates the problem \cite{gippius2004nonlinear}, necessitating the coupling of Maxwell's equations to the intensity-dependent response of the medium. At high polariton densities, nonlinear phenomena such as nonlinear resonances \cite{de2014control}, optical bistability, multistability, and self-focusing become prominent \cite{walker2015ultra}, requiring an extension of the TMM to incorporate these effects.

TMM offers clear advantages in modeling propagation of light through layered structures due to its simplicity and relatively low computing demands. However, it is known that its applicability may be limited due to inherent instability in higher dimensions~\cite{perez2015relations,luque2013symmetry}. The instability is usually related to complex eigenvalues of certain modes in the presence of strong absorption or transverse surface modes. In the case of an optical microcavity incorporating thin quantum wells such modes may become suppressed, leading to numerical stability.
Other numerical methods have also been employed to solve Maxwell’s equations~\cite{scarmozzino2000numerical}. For instance, the Finite Difference Time Domain (FDTD) method is a popular alternative that discretizes both space and time to directly simulate the evolution of electromagnetic fields \cite{kunz1993finite, taflove2005computational,dini2024nonlinear}. FDTD is particularly advantageous when dealing with complex geometries and inhomogeneous media, but it can be computationally expensive for large-scale systems or high-resolution simulations. Another approach is the finite element method (FEM), which offers flexibility in handling arbitrary geometries but may suffer from increased computational overhead, especially in multilayered systems \cite{reddy1993introduction, rao2010finite}. Sophisticated methods such as the semianalytical method of lines or admittance transfer method provide stable and efficient solvers of the linear eigenmode problem~\cite{pregla2008analysis,dems2005planewave}.

In this work, we focus on extending the TMM to include both system inhomogeneity in transverse dimensions and nonlinear exciton-polariton interactions. We note that the existing literature includes many works devoted to these aspects, both in the general context of nonlinear optics and in the specific case of exciton-polariton systems. For example, in Refs.~\cite{diao2007optical,novitsky2008bistable,entezar2013optical,mostafazadeh2019nonlinear,pakhomov2021modeling} nonlinear extensions of the TMM in one dimension were investigated. In the linear, homogeneous layers case, the finite beam size can be taken into account simply by considering the oblique incidence for each of the Fourier components of the input beam~\cite{born2013principles,kavokin2017microcavities,ivchenko1997resonance}. However, both in the cases of an inhomogeneous system and a nonlinear system, mixing of Fourier modes with different transverse components occurs, which increases the complexity of calculations  considerably~\cite{pilozzi2007effect}. A general theory of TMM in the linear, inhomogeneous case was developed in~\cite{loran2016transfer,loran2020transfer}, but its numerical stability was not investigated, which is a crucial point in the light of the limited stability of multidimensional TMM~\cite{perez2015relations,luque2013symmetry}. In the exciton-polariton case, a Green's function method capable of calculating TE polarized modes of linear systems with periodic modulation in the transverse direction was developed in~\cite{pilozzi2007effect}. However, to our best knowledge, TMM method has not been successfully used in the case of nonlinear systems beyond the one-dimensional case. Moreover, it is not clear whether TMM can be used to provide a stable, efficient numerical method in the case of more general inhomogeneous  systems and nonlinear systems.

In this work, we extend these works by demonstrating a stable, practical and resource-efficient implementation of TMM in a multidimensional, nonlinear exciton-polariton system. By incorporating the effects of nonlinearities into the TMM framework, we provide a more accurate representation of light-matter interactions in these systems. Through this extension, we aim to bridge the gap between linear methods and more computationally intensive approaches like FDTD and FEM. Moreover, it provides a simple and numerically stable algorithm for finding eigenmodes of Maxwell equations in polariton microcavities. Our results offer insights into the optical properties of polariton microcavities, paving the way for the development of novel quantum photonic technologies and nonlinear light sources.

The structure of our paper is as follows. In section II we provide an introduction to the TMM in three dimensions. In Sec.~III we generalize it to the case of inhomogeneous medium, deriving our variant of this method including mixing of Fourier modes. In Sec.~IV we focus on the implementation of a nonlinear quantum well layer and in Sec.~V we analyze boundary conditions and different polarizations.  This extension is presented in detail in Sections IV of our work. Additionally, the question of numerical instability when extending the TMM to more than one dimension is addressed in Section VI, where we predict the effects of self-focusing (lensing) and spontaneous symetry breaking in a system incorporating a thin quantum well. IN Sec.~VII we demonstrate the superiority of the TMM method by comparing execution time of our algorithm with other numerical methods for solving Maxwell's equations.

\section{Transfer Matrix Method in Three Dimensions}

The transfer matrix method is a powerful technique for describing the propagation of electromagnetic waves across a sequence of layers stacked along the $z$-axis. This method is particularly elegant and computationally efficient for analyzing multilayer systems, especially when the layers are homogeneous and isotropic. Below, we present a comprehensive derivation, based on \cite{kavokin2017microcavities}, of the transfer matrix for both transverse-electric (TE) and transverse-magnetic (TM) polarized electromagnetic fields, rooted in Maxwell's equations \cite{maxwell1865viii}.

Let us begin with Faraday's law and Ampère-Maxwell's law in the frequency domain
\begin{align}
    \nabla \times \vec{E} &= -i\omega \vec{B}, \label{eq:faraday} \\
    \nabla \times \vec{H} &= i\omega \vec{D}, \label{eq:ampere_maxwell}
\end{align}
where the material relations are $\vec{D} = \epsilon \vec{E}$ and $\vec{B} = \mu \vec{H}$. For a homogeneous, isotropic medium with refractive index $n$, we use $\epsilon = n^2 \epsilon_0$ and $\mu = \mu_0$. 

For a harmonic plane wave with wavevector $\vec{k} = (k_x, k_y, k_z)$ and magnitude $|\vec{k}| = n k_0 = \frac{n\omega}{c}$, the electric field can be expressed as
\begin{align}
    \vec{E}(\vec{r}) = \vec{E}_0 e^{i (\vec{k} \cdot \vec{r})},
\end{align}
and similarly for the magnetic field $\vec{B}$. The relationship between $k_z$ and $\vec{k}$ is given by
\begin{align}
    k_z = n k_0 \cos\theta, \label{eq:kz_relation}
\end{align}
where $\theta$ is the angle of incidence relative to the normal of the interface.

The evolution of the electric field is governed by the wave equation. For the transverse-electric (TE, or s-) polarized component in the $\tau$ direction (in the plane of incidence), the wave equation simplifies to
\begin{align}
    \frac{\partial^2 E_{\tau s}}{\partial z^2} + k_z^2 E_{\tau s} = 0. \label{eq:wave_eq_TE}
\end{align}
The general solution to this differential equation is a superposition of forward and backward propagating waves 
\begin{align} 
    E_{\tau s}(z) &= E_0^+ e^{i k_z z} + E_0^- e^{-i k_z z}, \label{eq:E_tau_solution} \\
    B_{\tau s}(z) &= \frac{k_z}{\omega} \left( E_0^+ e^{i k_z z} - E_0^- e^{-i k_z z} \right). \label{eq:B_tau_solution} 
\end{align} 
These expressions describe the electric and magnetic fields in terms of their forward ($E_0^+$) and backward ($E_0^-$) propagating wave components. These relationships can be expressed in matrix form to systematically relate the fields in terms of left- and right-propagating components. Introducing a change of basis matrix $C$, the fields amplitudes are expressed as 
\begin{align} 
    \begin{bmatrix} E \\ cB \end{bmatrix} = \begin{bmatrix} 1 & 1 \\ n \cos \theta & -n \cos \theta\end{bmatrix} \begin{bmatrix} E^+ \\ E^- \end{bmatrix} = C \begin{bmatrix} E^+ \\ E^- \end{bmatrix}, \label{eq:base_change} 
\end{align} 
where $E^+$ and $E^-$ represent the amplitudes of the right- and left-propagating electric fields, respectively.

This matrix formulation simplifies the treatment of boundary conditions and the construction of the overall transfer matrix for multilayer structures, as it provides a convenient framework to transition between field components and their respective propagation directions. The use of the change of basis matrix $C$ is central to connecting the field continuity equations across interfaces and analyzing the wave behavior in complex optical systems.

Consider a homogeneous layer with boundaries $z = 0$ and $z = d$. Specifically, at $z = 0$
\begin{align}
    E_{\tau s}(0) &= E_0^+ + E_0^-, \label{eq:boundary_E0} \\
    cB_{\tau s}(0) &= \frac{k_z}{k_0} \left( E_0^+ - E_0^- \right), \label{eq:boundary_B0}
\end{align}
and at $z = d$
\begin{align}
    E_{\tau s}(d) &= E_0^+ e^{i k_z d} + E_0^- e^{-i k_z d}, \label{eq:boundary_Ed} \\
    cB_{\tau s}(d) &= \frac{k_z}{k_0} \left( E_0^+ e^{i k_z d} - E_0^- e^{-i k_z d} \right). \label{eq:boundary_Bd}
\end{align}

These boundary conditions can be encapsulated in a matrix equation relating the fields at $z = d$ to those at $z = 0$
\begin{align}
    \begin{bmatrix}
        E_{\tau s}(d) \\ cB_{\tau s}(d)
    \end{bmatrix} &= \begin{bmatrix}
        \cos(k_z d) & \frac{i}{n \cos \theta} \sin(k_z d) \\
        i n \cos \theta \sin(k_z d) & \cos(k_z d)
    \end{bmatrix}
    \begin{bmatrix}
        E_{\tau s}(0) \\ cB_{\tau s}(0)
    \end{bmatrix} \nonumber\\
    &= T_s \begin{bmatrix}
        E_{\tau s}(0) \\ cB_{\tau s}(0)
    \end{bmatrix}, \label{eq:transfer_matrix_TE}
\end{align}
where the diagonal elements $\cos(k_z d)$ account for the phase evolution due to $e^{\pm i k_z d}$, and the off-diagonal terms represent the coupling between $E_\tau$ and $cB_\tau$ determined by the refractive index $n$ and the angle of incidence $\theta$.

Similarly, using the symmetry of the system of Eqs.~(\ref{eq:faraday}) and (\ref{eq:ampere_maxwell}), for the transverse-magnetic (TM, or p-) polarized field, the transfer matrix has a similar form in the basis $ \begin{bmatrix} cB_{\tau p} \\ E_{\tau p} \end{bmatrix} $. Following \cite{born2013principles}, the corresponding transfer matrix for the TM-polarized field is given by
\begin{align}
    T_p &= \begin{bmatrix}
        \cos(k_z d) & \frac{i n}{\cos \theta} \sin(k_z d) \\
        \frac{i \cos \theta}{n} \sin(k_z d) & \cos(k_z d)
    \end{bmatrix}. \label{eq:transfer_matrix_TM}
\end{align}
This matrix accounts for the phase accumulation and amplitude changes due to the refractive index and the angle of propagation, analogous to the TE case but adjusted for TM polarization.

To handle oblique angles of incidence and to simplify the analysis, it is advantageous to work in momentum space. By applying a Fourier transform in the $x$ and $y$ directions, the field in momentum space, $\widetilde{F}(k_x, k_y)$, is expressed as
\begin{align}
    \widetilde{F}(k_x, k_y) = \frac{1}{2 \pi} \int_{-\infty}^\infty \dd{x} \dd{y} F(x, y) \ee^{i (k_x x + k_y y)}. \label{eq:fourier_transform}
\end{align}
This transformation allows us to treat  wave propagation in terms of spatial frequencies $k_x$ and $k_y$, facilitating the analysis of structured media.

Given the wavelength $\lambda$ of the incoming electromagnetic field, the wavevector in a medium with refractive index $n$ is 
\begin{align} 
    \vec{k} = \left[ k_x, k_y, \sqrt{\frac{4\pi^2 n^2}{\lambda^2} - k_x^2 - k_y^2} \right], \label{eq:wavevector_general} 
\end{align}
The cosine of the angle of incidence with respect to the normal equals
\begin{align} 
    \cos \theta = \sqrt{1 - \frac{\lambda^2 (k_x^2 + k_y^2)}{4\pi^2 n^2}}. \label{eq:cos_theta_general}
\end{align}

The theoretical simplicity of the transfer matrix method arises from the ability to construct the transfer matrix for a multilayer semiconductor structure by sequentially left-multiplying the matrices corresponding to individual layers. Specifically, if a structure consists of multiple layers, each with its own transfer matrix $T_i$, the overall transfer matrix $T$ is obtained as the product
\begin{align} \label{eq:tttm}
    T = T_N T_{N-1} \cdots T_2 T_1,
\end{align}
where $N$ is the number of layers. This feature allows for an elegant and efficient analysis of complex multilayer systems. At each boundary between layers, the continuity of the electric and magnetic fields ensures that the fields can be seamlessly connected across interfaces. By expressing the fields in terms of the left- and right-propagating components and applying the transfer matrices derived above, one can solve for the transmission and reflection coefficients of the entire multilayer structure.

\section{Transfer Matrix Method in inhomogenous medium}
To model the behavior of electromagnetic waves in an inhomogeneous medium, we begin by deriving the wave equation in a medium with spatially varying permittivity $\epsilon(\vec{r})$. For simplicity, we consider the case where the polarization (specifically, TE polarization with the electric field polarized in the $x$-direction) is preserved during scattering. This approximation holds as long as variations in permittivity occur on spatial scales much larger than the wavelength. The Maxwell equation \cite{maxwell1865viii} describing the rotation of the electric field $\vec{E}$ is
\begin{align}
    \nabla \times \vec{E} = -\frac{\partial \vec{B}}{\partial t}. \label{Maxwell-Faraday}
\end{align}
Taking the curl of both sides of Eq.~\eqref{Maxwell-Faraday}, we obtain
\begin{align}
    \nabla \times (\nabla \times \vec{E}) = -\frac{\partial}{\partial t} (\nabla \times \vec{B}). \label{Curl-Curl-E}
\end{align}
Using Ampère-Maxwell's law, $\nabla \times \vec{H} = \frac{\partial \vec{D}}{\partial t}$, and assuming $\vec{D} =  \epsilon_0 \epsilon(\vec{r}) \vec{E}$ and $\vec{B} = \mu_0 \vec{H}$, Eq.~\eqref{Curl-Curl-E} simplifies to
\begin{align}
    \nabla \times (\nabla \times \vec{E}) = - \frac{\epsilon(\vec{r})}{c^2} \frac{\partial^2 \vec{E}}{\partial t^2}, \label{Wave-Operator}
\end{align}
where $c = \frac{1}{\sqrt{\mu_0 \epsilon_0}}$ is the speed of light in vacuum. Assuming that the electric field is transverse (i.e., $\nabla \cdot \vec{E} = 0$), we can use the vector identity $\nabla \times (\nabla \times \vec{E}) = \nabla (\nabla \cdot \vec{E}) - \nabla^2 \vec{E} = -\nabla^2 \vec{E}$. This leads to the wave equation
\begin{align}
    \nabla^2 \vec{E} = \frac{\epsilon(\vec{r})}{c^2} \frac{\partial^2 \vec{E}}{\partial t^2}, \label{Wave-Equation}
\end{align}
For a monochromatic field with angular frequency $\omega$, we substitute the time dependence $\vec{E}(\vec{r}, t) = \vec{E}(\vec{r}) e^{i \omega t}$ into Eq.~\eqref{Wave-Equation} to obtain the Helmholtz equation
\begin{align}
    \nabla^2 \vec{E} + \frac{\omega^2}{c^2} \epsilon(\vec{r}) \vec{E} = 0. \label{Helmholtz-Equation}
\end{align}

To solve Eq.~\eqref{Helmholtz-Equation} for layered media, we perform a Fourier transform in the $x$ and $y$ directions while keeping the $z$-direction in real space. This Fourier transform of the electric field is defined as 
\begin{align}
    \widetilde{\vec{E}}(k_x, k_y, z) = \frac{1}{2\pi} \int \vec{E}(x, y, z) e^{i (k_x x + k_y y)} \, dx \, dy. \label{Fourier-E}
\end{align}
Similarly, we apply the Fourier transform to the permittivity $\epsilon(x, y, z)$ to obtain $\widetilde{\epsilon}(k_x, k_y, z)$. Substituting these transforms into the Helmholtz equation and applying the convolution theorem for Fourier transforms, we obtain
\begin{align}
    \frac{\partial^2 \widetilde{\vec{E}}}{\partial z^2} - (k_x^2 + k_y^2) \widetilde{\vec{E}} = -\frac{\omega^2}{c^2} \left( \widetilde{\epsilon} \star \widetilde{\vec{E}} \right), \label{Helmholtz-Fourier}
\end{align}
where $\widetilde{\epsilon} \star \widetilde{\vec{E}}$ represents the convolution of the permittivity and the electric field in momentum space,
\begin{align}
    \widetilde{\epsilon} \star \widetilde{\vec{E}} = \int \widetilde{\epsilon}(k_x', k_y', z) \widetilde{\vec{E}}(k_x - k_x', k_y - k_y', z) \, dk_x' \, dk_y'. \label{Convolution-Term}
\end{align}
This convolution term accounts for both the spatial distribution of permittivity and potential nonlinear refractive index changes in the medium, particularly in layered and inhomogeneous materials where the system's response depends on the field distribution.

Next, we express Eq.~\eqref{Helmholtz-Fourier} in a more compact operator form. Let $\omega = k_0 c$, where $k_0 = \omega / c$ is the free-space wavenumber. The equation becomes
\begin{align} \label{eq:conv_differential}
    \frac{\partial^2 \widetilde{\vec{E}}}{\partial z^2} + \mathcal{K}_\epsilon \widetilde{\vec{E}} &= 0, \\
    \mathcal{K}_\epsilon(k_x, k_y, z) &= - (k_x^2 + k_y^2) + k_0^2 \widetilde{\epsilon}(k_x, k_y, z) \star. \nonumber
\end{align}
The operator $\mathcal{K}_\epsilon$ in Eq.~\eqref{eq:conv_differential} can be represented in matrix form as follows. Considering discrete indices $i$ and $j$ corresponding to discretized $k_x$ and $k_y$ values, respectively, the matrix elements of $\mathcal{K}_\epsilon$ are given by
\begin{align}
    \mathcal{K}_\epsilon[i, i', j, j'] (z)&= k_0^2 \widetilde{\epsilon}\left(k_x[i] - k_x[i'], k_y[j] - k_y[j'], z\right) \nonumber \\
    &\quad - \delta_{i, i'} \delta_{j, j'} \left(k_x[i]^2 + k_y[j]^2\right). \label{K-epsilon-Matrix-Discrete}
\end{align}
Here, $\widetilde{\epsilon}(k_x, k_y, z)$ is the Fourier-transformed permittivity, and the convolution is captured by the term $\widetilde{\epsilon}\left(k_x[i] - k_x[i'], k_y[j] - k_y[j'], z\right)$. The Kronecker deltas ensure that the transverse wavenumber terms are only subtracted on the diagonal of the matrix.

The general solution to Eq.~\eqref{eq:conv_differential} involves forward and backward propagating waves, represented in terms of the eigenvalues and eigenvectors of $\mathcal{K}_\epsilon$. Under assumption that $\widetilde{\epsilon}(k_x,k_y,z) = \widetilde{\epsilon}(k_x,k_y,z+\Delta z)$ for small $\Delta z$, the electric and magnetic fields are expressed as (cf. Eqs.~(\ref{eq:E_tau_solution}) and (\ref{eq:B_tau_solution}))
\begin{align}
    \vec{E}(z+\Delta z) &= e^{i \Delta z \sqrt{\mathcal{K}_\epsilon}} \vec{E}^+(z) + e^{-i \Delta z \sqrt{\mathcal{K}_\epsilon}} \vec{E}^-(z), \label{Solution-E} \\
    c\vec{B}(z+\Delta z) &= \frac{1}{k_0} \sqrt{\mathcal{K}_\epsilon} \left(e^{i \Delta z \sqrt{\mathcal{K}_\epsilon}} \vec{E}^+(z) - e^{-i \Delta z \sqrt{\mathcal{K}_\epsilon}} \vec{E}^-(z)\right), \label{Solution-B}
\end{align}
where the operator $\sqrt{\mathcal{K}_\epsilon}$ is defined via the eigen-decomposition
\begin{align}
    \sqrt{\mathcal{K}_\epsilon} = V_\epsilon \sqrt{\Lambda_\epsilon} V_\epsilon^{-1},
\end{align}
with $V_\epsilon$ containing the eigenvectors and $\Lambda_\epsilon$ denoting the diagonal matrix of eigenvalues of $\mathcal{K}_\epsilon$.

The transfer matrix $T_N(k_x, k_y)$, which propagates the fields from position $z$ to $z + \Delta z$, is then given by
\begin{align}
    T_\text{N}(k_x, k_y) = \begin{bmatrix}
        \cos(\Delta z \sqrt{\mathcal{K}_\epsilon}) & i k_0 \sin(\Delta z \sqrt{\mathcal{K}_\epsilon}) \sqrt{\mathcal{K}_\epsilon}^{-1} \\
        \frac{i}{k_0} \sin(\Delta z \sqrt{\mathcal{K}_\epsilon}) \sqrt{\mathcal{K}_\epsilon} & \cos(\Delta z \sqrt{\mathcal{K}_\epsilon})
    \end{bmatrix}, \label{Transfer-Matrix}
\end{align}
where $\sqrt{\mathcal{K}_\epsilon}^{-1}$ serves as the deconvolution operator, satisfying $\sqrt{\mathcal{K}_\epsilon}^{-1} \sqrt{\mathcal{K}_\epsilon} = \mathbb{I}$. This relation holds provided that all eigenvalues of $\mathcal{K}_\epsilon$ are non-zero (ie. $\mathcal{K}_\epsilon$ is invertible), a condition typically met under physical circumstances.

For small $\Delta z$, the transfer matrix can be approximated by expanding the trigonometric functions to first order
\begin{align}
    T_\text{N}(k_x, k_y) \approx \begin{bmatrix}
        1 & i k_0 \Delta z \\
        \frac{i}{k_0} \Delta z \, \mathcal{K}_\epsilon & 1
    \end{bmatrix}. \label{Transfer-Matrix-Approx}
\end{align}
This linear approximation simplifies the computation for thin layers, enabling iterative propagation through slices where changes in permittivity along the $z$-direction are negligible.

\section{Transfer Matrix Method for the exciton-polariton quantum well}
Now, we turn to the TMM as applied to the exciton-polariton quantum well (QW). For a QW in the one-dimensional case, the transfer matrix in the $\begin{bmatrix} E \\ cB \end{bmatrix}$ basis is given by~\cite{kavokin2017microcavities} 
\begin{align} \label{linear-tmm-qw} 
    T_{\text{QW}} = \begin{bmatrix} 1 & 0 \\ 2 n \frac{r_{\text{QW}}}{t_{\text{QW}}} & 1 \end{bmatrix}, 
\end{align} 
where $n$ is the refractive index of the medium, and $r_{\text{QW}}$ and $t_{\text{QW}}$ are the reflection and transmission coefficients in the case of a QW layer in vacuum. It can be checked that this form of transfer matrix is actually valid in general for a thin slice of any material of much smaller thickness than $\lambda$. However, in the case of QW, these coefficients depend nonlinearly on the exciton density, which is related to the local electric field intensity. We assume that the reflection and transmission coefficients are given by~\cite{kavokin2017microcavities} 
\begin{align} 
    r_{\text{QW}}(x, y; \omega) &= \frac{i \Gamma_0}{\omega_{\text{QW}} - \omega - g n_{\rm ex} - i \Gamma_0}, \label{eq:r_qw} \\
    t_{\text{QW}}(x, y; \omega) &= 1 + r_{\text{QW}}(x, y; \omega), \label{eq:t_qw}
\end{align} 
where $\omega_{\text{QW}}$ is the resonant frequency of the polariton microcavity, $\Gamma_0$ is the exciton radiative broadening, and $g$ represents the strength of the  exciton-exciton interaction. The parameter $n_{\rm ex}$ denotes the exciton density~\cite{estrecho2019direct}. In a stationary solution, $n_{\rm ex}(x,y) \propto |E(x,y,z=z_{\rm QW})|^2$, indicating that the field intensity within the quantum well is directly related to the exciton density. We will denote the detuning by $\Delta$, defined for linear systems as $\Delta = \omega_{\text{QW}} - \omega$ and for nonlinear systems as $\Delta = \omega_{\text{QW}} - \omega - g n_{\rm ex}$. This parameter will play a key role in the description of resonance behavior and interaction dynamics within the system.

To solve for the full system at arbitrary incidence, we need to consider Fourier transforms of the fields. The TMM, in the case of exciton-polaritons, can be extended to handle the nonlinear effects, as described by Eq.\eqref{Transfer-Matrix-Approx} and Eq.\eqref{eq:conv_differential}. We assume that the quantum well corresponds to the limit $\Delta z \to 0$, but care must be taken when taking this limit. As we wish to have the situation when the QW layer results in a finite phase shift, the factor $\Delta z\, \widetilde{\epsilon}$ should converge to a finite value
\begin{align}
    \Delta z \, \mathcal{K}_\epsilon &= - \Delta z (k_x^2 + k_y^2) + \Delta z k_0^2 \widetilde{\epsilon}(k_x, k_y, z) \star \\
    &\to k_0 \widetilde{\epsilon}_{\rm eff}(k_x, k_y, z) \star,
\end{align}
then Eq.\eqref{Transfer-Matrix-Approx} becomes 
\begin{align} \label{eq:tmm_qw}
    T_\text{QW}(k_x, k_y) = \begin{bmatrix} 1 & 0 \\ i \widetilde{\epsilon}_{\rm eff} \star & 1 \end{bmatrix}.
\end{align}
Compared with Eq.~\eqref{linear-tmm-qw}, the effective permittivity of the non-linear quantum well is 
\begin{align}
    i\widetilde{\epsilon}_{\rm eff}(k_x,k_y)= 2in \Gamma_0 \; \int_{-\infty}^\infty \int_{-\infty}^\infty \frac{\dd x \dd y \ee^{ik_x x + ik_y y}}{\omega_\text{QW} - \omega - g|E(x,y)|^2}.
\end{align}
In our model, we assume that the nonlocal dielectric response (i.e., the interaction with excitons located at different spatial positions and its resulting impact on the polarization) is negligible, given that the characteristic spatial extent (Bohr radius) of excitons is much smaller than the optical wavelength considered. Physically, this implies that the exciton-exciton interaction predominantly occurs locally, and thus the polarization at any given point is influenced primarily by the local electric field intensity.

Moreover, we have neglected the dispersion of the dielectric constant as compared to the nonlinear interaction term. Therefore, in our calculations, the effective permittivity does not explicitly depend on the wavevector. 
However, extending our formalism to explicitly incorporate spatial dispersion 
is straightforward. This can be achieved by redefining the detuning energy in the denominator
\begin{align}
   \Delta(\omega, k) = \omega_\text{QW}  + \frac{\hbar k^2}{2m}- \omega - g|E(x,y)|^2,
\end{align}
where $k$ is the wavevector and $m$ is the exciton translational mass. Finally, for a comprehensive theoretical description that accounts for decoherence mechanisms, one could additionally incorporate the homogeneous broadening term, $i\gamma$, into the dielectric function, providing a more realistic and experimentally relevant simulation of exciton-polariton systems.

The boundary conditions that need to be satisfied are analyzed in the linear regime in the next section, while for the non-linear QW case they are discussed in \hyperref[sec:app_a]{Appendix A}.

\section{Boundary Conditions}
To analyze wave propagation through the optical structure, we consider an incoming right-propagating (positive $k_z$) electric field \(E_{\rm in}^+(x, y)\), composed of plane waves with positive \(k_z\) components, incident from the left. The structure is oriented parallel to the \(xy\)-plane, and for simplicity, we assume that no left-propagating (negative \(k_z\)) electric field component, \(E_{\rm out}^-(x, y)\), is incident from the right side. This assumption provides a well-defined boundary condition, simplifying the determination of the field distributions within the structure. The boundary conditions and their derivation in the case of a nonlinear quantum well (QW) are further detailed in \hyperref[sec:app_a]{Appendix A}. 

\begin{figure}[hbtp]
    \centering
    \includegraphics[width=0.48\textwidth]{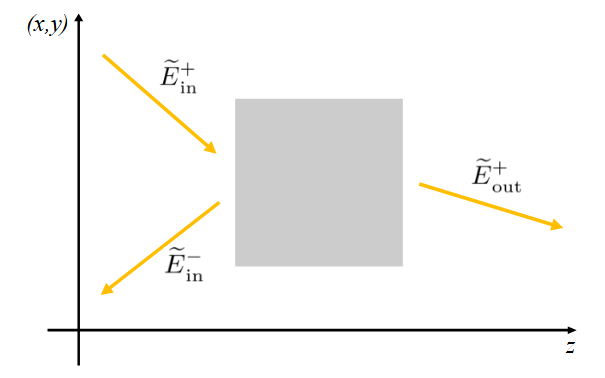}
    \caption{\justifying \fontsize{9pt}{11pt}\selectfont Schematic representation of wave propagation through the optical system. The structure is illuminated by an incoming electric field consisting of right- (positive $k_z$) and left-propagating (negative $k_z$) components, \(\widetilde{E}_{\text{in}}^+\) and \(\widetilde{E}_{\text{in}}^-\), respectively. After interacting with the system, the transmitted field \(\widetilde{E}_{\text{out}}^+\) exits on the right. The grey rectangle represents the structure.}
    \label{fig:wave_propagation}
\end{figure}

A schematic representation of the wave propagation scenario is illustrated in Fig.~\ref{fig:wave_propagation}. The incoming field consists of both a right-propagating (positive $k_z$) component, \(\widetilde{E}_{\text{in}}^+\), and a left-propagating (negative $k_z$) component, \(\widetilde{E}_{\text{in}}^-\), which interact with the structure. The transmitted field, \(\widetilde{E}_{\text{out}}^+\), emerges after propagation through the optical system.

\subsection{Polarized Right-Propagating Input Field}
We now focus on the incoming electromagnetic field. We determine the electric and magnetic components of $E^+_{\rm in}(x,y)$ for each of orthogonal polarizations. For simplicity, let us assume that the boundary conditions for the incoming right-propagating (positive $k_z$) field $ E^+_{\rm in} $ are specified by amplitude of its components at $z=0$ in function of the $ x $ and $ y $ variables. Since the electric field is perpendicular to the wavevector, the vector representation of the right-propagating electric field in vacuum can be expressed as
\begin{align}
    \vec{\widetilde{E}_{\rm in}}^+(k_x,k_y) = \begin{bmatrix}
        \widetilde{E}_x \\ \widetilde{E}_y\\ -\frac{k_x \widetilde{E}_x + k_y \widetilde{E}_y}{\sqrt{\frac{4\pi^2}{\lambda^2} - k_x^2 - k_y^2}}
    \end{bmatrix}.
\end{align}
On the other hand, the normalized basis vectors corresponding to the $ TE $ (transverse electric) and $ TM $ (transverse magnetic) polarizations for the right-propagating field are given as
\begin{align}
    \vec{e}_{s+} &= \frac{\vec{k} \times \vec{e}_z}{|\vec{k} \times \vec{e}_z|} = \frac{1}{\sqrt{k_x^2 +k_y^2}} \begin{bmatrix}
        k_y \\ -k_x \\ 0
    \end{bmatrix},  \label{eq:vers.s+} \\
    \vec{e}_{p+} &= \frac{\vec{k} \times (\vec{k} \times \vec{e}_z)}{|\vec{k} \times (\vec{k} \times \vec{e}_z)|} = \frac{1}{k_0\sqrt{(k_x^2 +k_y^2)}} \begin{bmatrix}
        k_x k_z \\ k_y k_z \\ -k_x^2 - k_y^2
    \end{bmatrix}, \label{eq:vers.p+}
\end{align}
where $k_0 = \frac{2 \pi}{\lambda}$ and $k_z $ is determined from the dispersion relation as described in Eq.~\eqref{eq:wavevector_general}. These polarization vectors serve as the orthonormal basis for the electric field decomposition.

The reflected left-propagating field $ \vec{\widetilde{E}}^- $ is not yet known. To find a solution, we start by projecting the incoming $ \vec{\widetilde{E}_{\rm in}}^+ $ field onto each of the polarization basis vectors. The transfer matrix for each layer is expressed in the $ \begin{bmatrix} \widetilde{E}_{\tau s/p} \\ c\widetilde{B}_{\tau s/p} \end{bmatrix} $ basis. These matrices are then multiplied layer by layer, followed by a basis transformation to the $ \begin{bmatrix} \widetilde{E}^+_{s/p} \\ \widetilde{E}^-_{s/p} \end{bmatrix} $ basis, leading to the transfer matrix for the entire multilayer structure in the $TE$ or $TM$ polarization
\begin{align}
    M_{s/p} = C^{-1} T_{s/p} C,
\end{align}
where $C$ and $T_{s/p}$ are given, respectively, by Eq.~\eqref{eq:base_change} and Eq.~\eqref{eq:transfer_matrix_TE} or Eq.~\eqref{eq:transfer_matrix_TM} as functions of $(k_x, k_y)$. This change of basis is essential to impose the physical boundary conditions at the right end of the structure, which is described by
\begin{align} \label{0cond}
    M_{s/p} \begin{bmatrix} \widetilde{E}^+_{s/p, {\rm in}} \\ \widetilde{E}^-_{s/p, {\rm in}} \end{bmatrix} = \begin{bmatrix} \widetilde{E}^+_{s/p, {\rm out}} \\ 0 \end{bmatrix},
\end{align}
which means that there is no left-propagating wave incident on the structure from the right. By solving this equation, the reflected field $ \widetilde{E}^- $ can be determined. Once the reflected field $ \widetilde{E}^- $ is obtained, the total electric field $ \widetilde{E} $ is reconstructed by summing the contributions from all polarization components. The resulting field is given by
\begin{align}
    \vec{E} = \widetilde{E}^+_{s} \vec{e}_{s+} + \widetilde{E}^-_{s} \vec{e}_{s-} + \widetilde{E}^+_{p} \vec{e}_{p+} + \widetilde{E}^-_{p} \vec{e}_{p-},
\end{align}
where the polarization vectors for the left-propagating field, with wavevector $ \vec{k} = (k_x, k_y, -k_z) $, are as follows
\begin{align}
    \vec{e}_{s-} &= \vec{e}_{s+},  \label{eq:vers.s-} \\
    \vec{e}_{p-} &= \frac{1}{k_0\sqrt{(k_x^2 +k_y^2)}} \begin{bmatrix}
        -k_x k_z \\ -k_y k_z \\ -k_x^2 - k_y^2
    \end{bmatrix}. \label{eq:vers.p-}
\end{align}
This final representation combines the contributions from both $ TE $ and $ TM $ polarizations for both directions of propagation, ensuring consistency with the physical boundary conditions and the properties of electromagnetic waves in multilayer systems.

\subsection{Polarized Input Field}
A simplified approach assumes that the incoming field has a specific polarization direction, the same for both the left- and right-propagating fields, for example, $\vec{e}_y$. For a given momentum $k$, once the direction of the electric field unit vector, $\vec{e}_E$, is known, the unit vector for the magnetic field can be determined as $\vec{e}_B = \frac{\vec{k} \times \vec{e}_E}{|\vec{k} \times \vec{e}_E|}$. With this information, it becomes possible to project the incoming electric and magnetic fields onto 
$s$ and $p$ polarizations. This projection is achieved using the projection matrix $P$
\begin{align}
    \begin{bmatrix}
        \widetilde{E}_{\tau s} \\
        c\widetilde{B}_{\tau s} \\
        c\widetilde{B}_{\tau p} \\
        \widetilde{E}_{\tau p}
    \end{bmatrix} &= \begin{bmatrix}
        \vec{e}_E \vec{e}_{\tau s} & 0 \\
        0 & \vec{e}_B \vec{e}_{\tau p} \\
        0 & \vec{e}_B \vec{e}_{\tau s}  \\
        \vec{e}_E \vec{e}_{\tau p} & 0
    \end{bmatrix} \begin{bmatrix}
        \widetilde{E} \\
        c\widetilde{B}
    \end{bmatrix} = P\begin{bmatrix}
        \widetilde{E} \\
        c\widetilde{B}
    \end{bmatrix}, \label{eq:polarizer}
\end{align}
where $\vec{e}_{\tau s} = \vec{e}_{s+}$ and $\vec{e}_{\tau p} = \vec{e}_{p_+} - (\vec{e}_{p_+} \vec{e}_z) \vec{e}_z$, resulting from the projection of the polarization versors onto the transverse plane. This framework allows for separating the contributions from each polarization in a systematic way. The transfer matrix for the entire structure, expressed in the $[\widetilde{E}_{\tau s}, c\widetilde{B}_{\tau s}, c\widetilde{B}_{\tau p}, \widetilde{E}_{\tau p}]^T$ basis, takes the form of a $4 \times 4 \times N_{k_x} \times N_{k_y}$ block matrix
\begin{align}
    T = \begin{bmatrix}
        \prod \limits_{j=1}^N T_{s,j}(k_x,k_y) & 0_{2\times2} \\
        0_{2\times2} & \prod \limits_{j=1}^N T_{p,j}(k_x,k_y)
    \end{bmatrix}. \label{eq:transferer}
\end{align}
This procedure separates the contributions of $s$-polarized and $p$-polarized components through diagonal block matrices, where $T_{s,j}(k_x, k_y)$ and $T_{p,j}(k_x, k_y)$ represent the transfer matrices for each layer $ j $ of the structure. Finally, the transfer matrix in the $\begin{bmatrix} \widetilde{E}^+ \\ \widetilde{E}^- \end{bmatrix}$ basis is obtained through a sequence of matrix multiplications
\begin{align}
    M = C^{-1} \cdot P^{-1} \cdot T \cdot P \cdot C,
\end{align}
where $A^{-1}$ denotes the pseudoinverse of matrix $A$ and matrices $C,P,T$ are described respectively by Eqs. \eqref{eq:base_change}, \eqref{eq:polarizer} and \eqref{eq:transferer}. This process transforms the $T$ matrix into a compact $2 \times 2 \times N_{k_x} \times N_{k_y}$ matrix
\begin{align}
    M = \begin{bmatrix}
        m_{11}(k_x,k_y) & m_{12}(k_x,k_y) \\ 
        m_{21}(k_x,k_y) & m_{22}(k_x,k_y)
    \end{bmatrix}.
\end{align}
The boundary condition that the left-propagating electric field vanishes at the right side of the structure provides the following relations
\begin{align}
    \widetilde{E}^-(k_x,k_y) &= - \frac{m_{21}(k_x,k_y)}{m_{22}(k_x,k_y)} \widetilde{E}_{\rm in}^+(k_x,k_y),\\
    \widetilde{E}^+_\mathrm{out}(k_x,k_y) &= \left(m_{11} - \frac{ m_{12} m_{21}}{m_{22}}\right)(k_x,k_y) \, \widetilde{E}_{\rm in}^+(k_x,k_y).
\end{align}
These expressions allow for the computation of the outgoing electric field, $\widetilde{E}_\mathrm{out}(k_x, k_y)$, as a function of the incoming electric field, $\widetilde{E}_{\rm in}^+(k_x, k_y)$. The dependence on the transfer matrix components makes this approach both flexible and powerful, enabling detailed analysis of wave propagation through complex structures.

\section{Numerical Simulations for Nonlinear Structures}
In this section, we present two numerical results from our simulations. The first demonstrates the self-focusing of a freely propagating Gaussian beam through the nonlinear quantum well. The second is an exciton-polariton microcavity, represented by a Bragg mirror microcavity with a nonlinear quantum well, where spontaneous symmetry breaking occurs as a result of interference of two symmetric Gaussian beams with opposite transverse wavevectors. Additional validation of TMM simulations using theoretical and numerical methods is discussed in the \hyperref[sec:app_a]{Appendix B}. 

\subsection{Self-Focusing in Nonlinear Quantum Well}
\begin{figure}[hbtp]
    \includegraphics[width=0.49\textwidth]{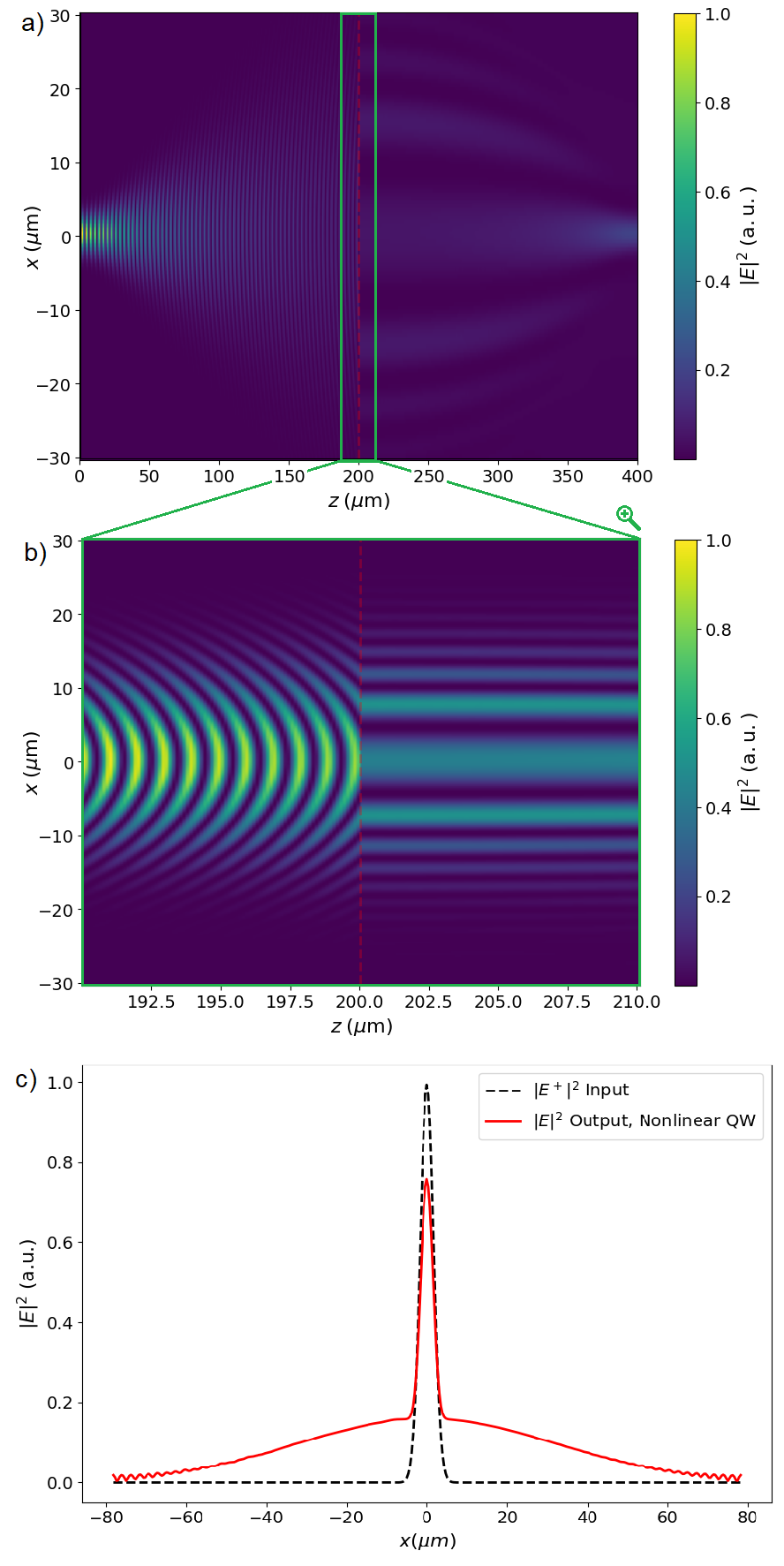}
    \caption{\justifying \fontsize{9pt}{11pt}\selectfont Self-focusing of a single Gaussian beam (width \(w_0=1.7\sqrt{2}\, \mu \text{m}\)) induced by the nonlinearity of an exciton quantum well (QW), represented by a dashed red line. a) Overview of the air-QW-air structure, showing beam propagation through the air regions (length \(d_\text{air}=200\mu\text{m}\)) and self-focusing after transmission through the infinitesimally thin nonlinear QW (dashed red line). b) Close-up view near the QW, illustrating the continuity of the electric field at the boundary. c) Comparison of the incoming and outgoing electric fields, showing the self-focused output mode alongside the standard spreading of the Gaussian input, after propagation through a QW-lens with a highly intensity-dependent refractive index.}
    \label{fig:self_focusing}
\end{figure}

Self-focusing effects are examined using a single Gaussian beam propagating through the air-QW-air structure, as shown in Fig.~\ref{fig:self_focusing}. Nonlinear optics literature describes self-focusing as a phenomenon in which the refractive index modification, induced by the intensity of the propagating field, leads to beam narrowing \cite{kivshar2003optical}. A comprehensive review of experimental realizations of self-focusing is provided in \cite{boyd2009self}, focusing on diverse media rather than semiconductor structures specifically. These experiments demonstrate that intense laser beams propagating through nonlinear media can exhibit self-focusing, leading to filamentation effects, with recent studies emphasizing femtosecond pulse-induced filaments in air. A concise overview of self-action effects arising from intensity-dependent refractive indices can also be found in \cite{boyd2008nonlinear}. Numerous experimental demonstrations exist, including a particularly clear example of self-focusing through photopolymerization shown both theoretically and experimentally in \cite{kewitsch1996self}.

From a physical standpoint, self-focusing occurs when the optical Kerr effect induces a spatial variation of the refractive index that mimics the action of a focusing lens. This leads to a nonlinear phase shift across the transverse profile of the beam. Specifically, the central region of a Gaussian beam, where intensity is highest, experiences a larger refractive index than the edges, resulting in wavefront curvature toward the axis. The phase front bends inward, converging the beam and leading to spatial compression in the transverse direction. Here, we are interested in the distribution of electromagnetic field interacting with a thin nonlinear QW and the possible emergence of self-focusing in this kind of structure.

In the simulated propagation, shown in Fig.~\ref{fig:self_focusing}.a), the nonlinear response of the QW induces self-focusing, causing the beam to narrow after transmission through the QW, accompanied by interference patterns. A close-up view in Fig.~\ref{fig:self_focusing}.b) reveals the continuity of the electric field at the boundary. The structure consists of two regions of free propagation through air with $n_\text{air}=1$ and a length of $d_\text{air}=200\mu\text{m}$, separated by a thin QW layer of exciton-polaritons, described by Eqs.~\eqref{eq:tmm_qw}, \eqref{eq:r_qw}, and \eqref{eq:t_qw}. We use an iterative approach to simulate the nonlinear effects: in each step, the electric field at the QW interface is calculated, using the intensity value from the previous iteration as input for the nonlinear permittivity equation.

The self-focusing effect can be understood as transmission through a lens with an intensity-dependent refractive index. In the first-order approximation in the light intensity $|E_\text{QW}|^2$, the refractive index is taken as 
\begin{align} 
    n(x,y) = n_0 + n_2 |E_\text{QW}(x,y)|^2, 
\end{align} 
where $n_0$ and $n_2$ are the low-intensity refractive index and the second-order nonlinear refractive index, respectively. After propagating through the QW-lens, the incoming field is partially reflected, resulting in an interference pattern on the left side of the air-QW-air structure. Simultaneously, the transmitted light, partially focused, follows the standard dispersion relation, resulting in a bimodal intensity distribution with a broad Gaussian profile with a prominent central peak (indicative of self-focusing). This behavior is illustrated in Fig.~\ref{fig:self_focusing}c), where the incoming and outgoing fields are compared. This kind of bimodal distribution is reminiscent of Bose-Einstein condensation which can occur in nonlinear optical materials~\cite{Chiocchetta_2016,Krupa_Classical_Condensation}

This idealized model captures the essential physics of nonlinear beam shaping. The narrow QW region introduces an abrupt, localized phase shift due to the nonlinear index perturbation, which can be approximated in the thin-lens limit. The effective focal length of such a Kerr lens is inversely proportional to the input intensity, and the focusing becomes stronger with increasing intensity, as observed in the simulation.

\subsection{Exciton-Polariton Microcavity}
\begin{widetext}

\begin{figure}[H]
    \centering
    \includegraphics[width=\textwidth]{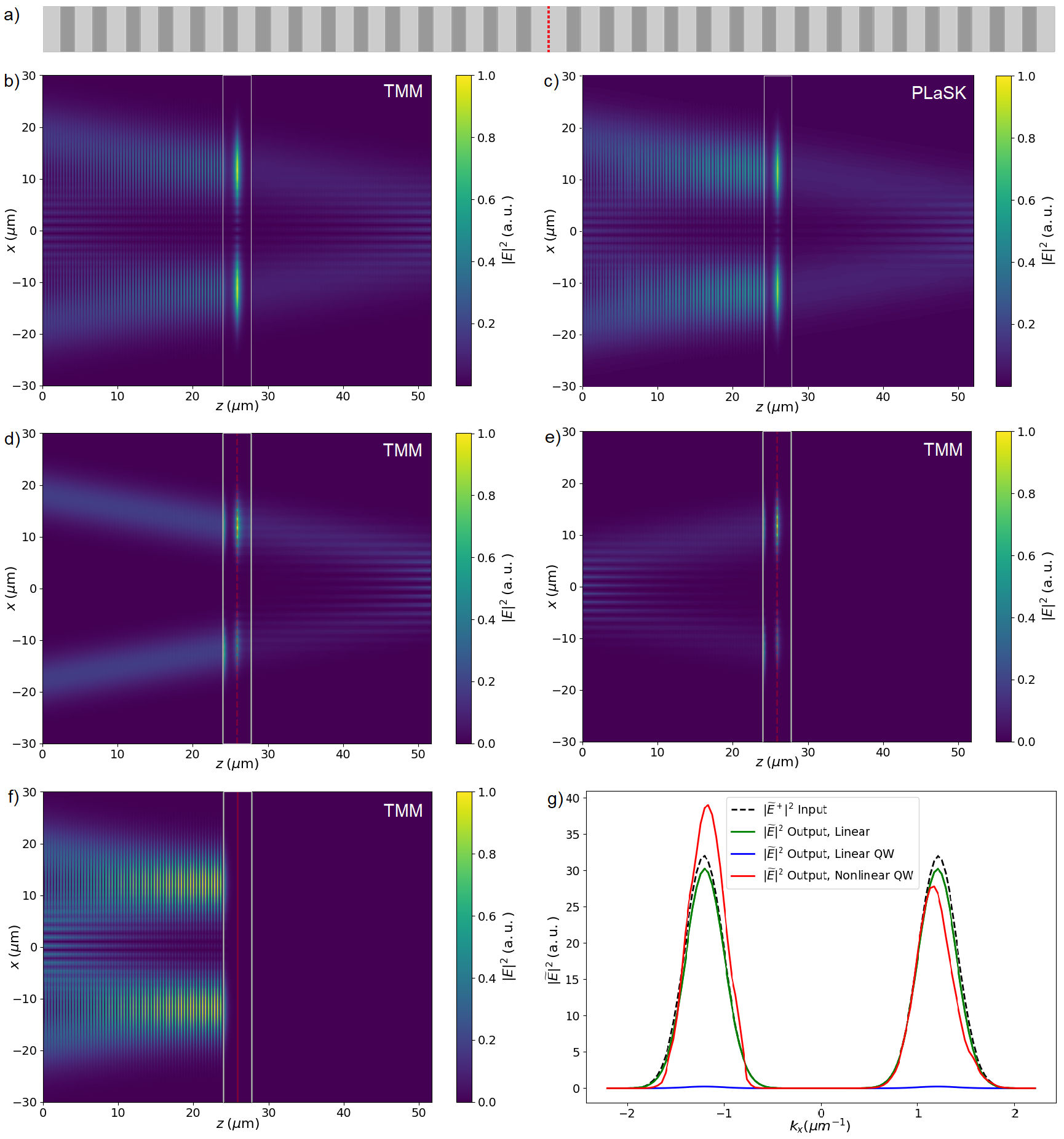}
    \caption{\justifying \fontsize{9pt}{11pt}\selectfont Comparison of field propagation and intensity distribution in a microcavity formed by two Bragg mirrors, each consisting of 15 repeating periods of alternating layers: light gray (\(n_a = 3, d_a = 0.065 \,\mu\text{m}\)) and dark gray (\(n_b = 3.5, d_b = 0.0577 \,\mu\text{m}\)), separated by a half-wavelength cavity (\(n_a, d = 2d_a\)), with or without a quantum well (QW) layer (dashed red line) at the center. The structure is resonant at \(\lambda = 4d_A n_A\) and embedded in air, with air layers \(24 \,\mu\text{m}\) thick before and after the microcavity to simulate free propagation. In all simulations, a Gaussian beam with a waist \(w_0 = 6\sqrt{2} \,\mu\text{m}\) is used, consisting of two beams separated by \(36 \,\mu\text{m}\), with a wavelength of \(\lambda = 780\) nm and transverse wavevectors \(k_x = \pm 6\pi \times 10^5 \, \frac{1}{\mu\text{m}}\). 
    a) Schematic of the polariton microcavity.  
    b) Interference of two Gaussian beams with opposite transverse wavevectors in a microcavity without a QW. The maximum intensity occurs in the microcavity center. The gray rectangle represents the microcavity.  
    c) Field distribution for the case depicted in b), calculated using PLaSK as a benchmark.   
    d) Nonlinear QW (dashed red line) with spontaneous symmetry breaking. Only the right-propagating (transmitted) field component is shown. Symmetry breaking increases the intensity of the upper outgoing beam, despite the equal intensity of the incoming beams.  
    e) Same as in d), but showing only the left-propagating (reflected) field component. 
    f) Similar to b), but with a linear QW (straight red line) at the microcavity center. A small detuning between the exciton and photon energy results in resonant reflection. 
    g) Fourier transforms of incoming and outgoing fields, showing distinct wavevectors and intensity modifications caused by spontaneous symmetry breaking in the nonlinear case.}
    \label{fig:microcavity}
\end{figure}
\end{widetext}

Numerical simulations of polariton microcavities with nonlinear quantum wells (QWs) reveal the dynamics of field propagation, interference effects, and the onset of nonlinearity-induced spontaneous symmetry breaking. The microcavity, shown in Fig.~\ref{fig:microcavity}, consists of alternating Bragg mirror layers with refractive indices \(n_a = 3\) and \(n_b = 3.5\), similar to AlGaAs alloys, with thicknesses \(d_a = 0.065 \, \mu \text{m}\) and \(d_b = 0.0557 \, \mu \text{m}\). These parameters ensure resonance at \(\lambda = 780 \, \text{nm}\) under normal incidence, satisfying \(\lambda = 4n_a d_a = 4n_b d_b\) \cite{kavokin2017microcavities}. The microcavity, depicted in Fig.~\ref{fig:microcavity}a), is surrounded by air layers of thickness \(d_{\text{air}} = 24 \, \mu \text{m}\) to allow free propagation.

Simulations were conducted for both linear and nonlinear QWs, as shown in Fig.~\ref{fig:microcavity}. In the linear case, the detuning parameter is fixed at \(\Delta = \omega_\text{QW} - \omega = -100 \, \Gamma = -10\, \text{meV}\), where \(\Gamma = 100 \,\mu\text{eV}\) is the exciton radiative broadening. In the nonlinear case, as described by Eq.~\eqref{eq:r_qw}, the detuning energy \(\Delta\) dynamically depends on the local electric field \(E_{\text{QW}}\) within the QW, following \(\Delta = -100 \, \Gamma \, (1 - 10^3 |E_{\text{QW}}|^2)\). The self-consistent field distribution is obtained iteratively, where the electric field at each step influences the next iteration, continuing until convergence. This iterative procedure captures the complex nonlinear dynamics, occasionally leading to instabilities for strong nonlinear effects.

In the linear regime (Fig.~\ref{fig:microcavity}b), two Gaussian beams with opposite transverse wavevectors interfere within the air-microcavity-air structure. Without a QW, the maximum intensity is observed at the cavity center, consistent with the spatial profile of microcavity modes~\cite{kavokin2017microcavities}.

Introducing a linear QW modifies the field distribution. Due to resonant reflection at small detuning (relative to the light-matter coupling strength), a significant portion of the field is reflected, as shown in Fig.~\ref{fig:microcavity}f).

\clearpage

The inclusion of a nonlinear QW leads to spontaneous symmetry breaking (SSB), a characteristic effect of nonlinear interactions. The right-propagating field entering the microcavity from the left is initially symmetric, with equal intensity in both Gaussian beams (Fig.~\ref{fig:microcavity}d)). However, after interacting with the nonlinear QW, symmetry breaks, resulting in unequal transmitted beam intensities. A similar asymmetry appears in the left-propagating reflected field (Fig.~\ref{fig:microcavity}e)), demonstrating SSB in both transmitted and reflected components. For clarity in demonstrating symmetry breaking, we have separated the results of the same propagation scenario into two distinct plots: one containing only the right-propagating components and the other containing only the left-propagating components. In Fig.~\ref{fig:microcavity}d) and Fig.~\ref{fig:microcavity}e), the upper beam inside the microcavity (center region), as well as after transmission (d) or reflection (e), exhibits higher intensity due to symmetry breaking occurring during propagation through the microcavity with a non-linear QW. However, immediately before entering the microcavity (d), both the upper and lower beams have identical intensities. This indicates that  interaction between left and right modes occurs inside the nonlinear structure, leading to the transfer of polaritons from the lower to the upper mode.

Such interaction and the resulting symmetry breaking can be compared with analogous phenomena occurring in confined structures, such as Josephson junctions~\cite{rahmani2016polaritonic}. Experimentally, spontaneous symmetry breaking in exciton-polariton systems has been studied in \cite{abbarchi2013macroscopic}, which analyzes Josephson oscillations resulting from quantum tunneling between two coupled polariton micropillars. In this work, the asymmetry is interpreted in terms of macroscopic self-trapping phenomenon.

Fourier analysis of the fields, shown in Fig.~\ref{fig:microcavity}g), highlights the distinct transverse wavevectors of the incoming \(E^+\) field, which remain unchanged across all cases, while the outgoing \(E\) field intensities differ in the nonlinear regime due to symmetry breaking.

For the linear microcavity without a QW, the results align closely with those obtained using the Photonic Laser Simulation Kit (PLaSK), a software for solving Maxwell's equations in semiconductor structures~\cite{dems2013automated}. This agreement, shown in Fig.~\ref{fig:microcavity}b), validates the numerical approach used in this study, confirming its accuracy in modeling microcavity structures.

Our simulations illustrate that the transfer matrix method is a physically accurate tool for modeling the complex propagation of electromagnetic field in polariton microcavities and nonlinear media. The results highlight the critical role of nonlinearity in shaping field distributions, symmetry-breaking behavior, and self-focusing effects.

\section{Computational efficiency}
Various numerical methods have been developed for solving Maxwell's equations in optical systems. These methods differ in their computational efficiency and execution time scaling with the number of modes $N$ \cite{scarmozzino2000numerical}. The beam propagation method (BPM), commonly used in paraxial approximation, scales as $O(N^2)$, making it efficient for systems with moderate mode counts but less suitable for highly detailed simulations. The finite-difference time-domain (FDTD) method, while versatile and accurate, scales as $O(N^3)$, due to the necessity of resolving Maxwell's equations over a spatial grid and time-stepping through the system dynamics. Similarly, the method of lines (MoL), employed in the Photonic Laser Simulation Kit (PLaSK), also exhibits a time complexity of \(O(N^3)\). This approach reduces partial differential equations to a system of ordinary differential equations, requiring discretization across transverse dimensions and subsequent matrix operations, akin to the computational demands of the FDTD method.

In this work, we introduce a novel solution based on the transfer matrix method (TMM) for systems with nonlinear permittivity. Our approach leverages the computational efficiency of Python's fast Fourier transform (FFT), achieving a time complexity of $O(N \log N)$. This substantial improvement in efficiency makes the method particularly well-suited for optical systems where nonlinearities necessitate iterative calculations over transverse Fourier modes. 

\begin{figure}[hbtp]
    \centering
    \includegraphics[width=0.49\textwidth]{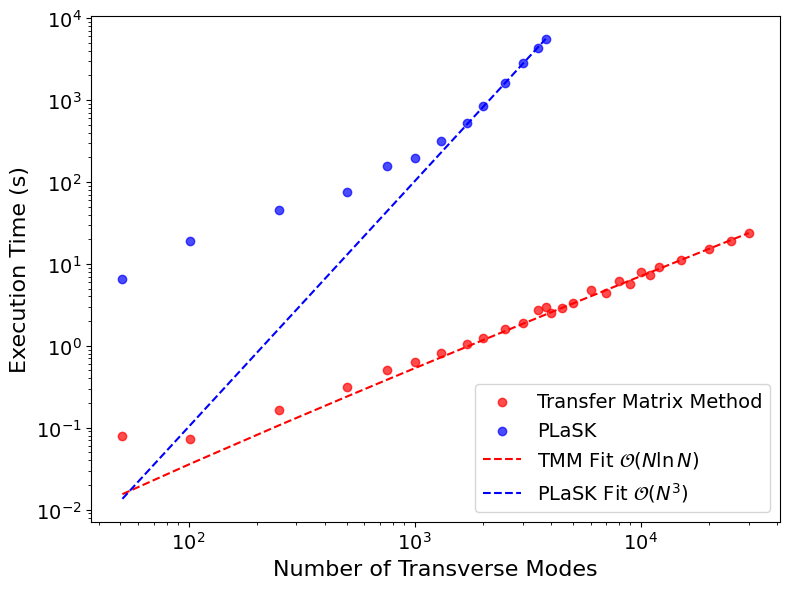}
    \caption{\justifying \fontsize{9pt}{11pt}\selectfont Comparison of computational efficiency between the Transfer Matrix Method (TMM) and Photonic Laser Simulation Kit (PLaSK) as a function of the number of Fourier modes. The plot illustrates the asymptotic behavior of \(\mathcal{O}(N^3)\) for PLaSK and \(\mathcal{O}(N \ln N)\) for TMM.}
    \label{fig:efficiency_comparison}
\end{figure}

In Fig.~\ref{fig:efficiency_comparison} we compare the execution time of our TMM-based method with the PLaSK software based on admittance transfer method on the same machine. The results demonstrate the superior scalability of TMM with number of modes, highlighting its potential for efficiently modeling large-scale optical systems with nonlinear effects.

\section{Conclusions}\label{Sec:Summary}
In this work, we extended the Transfer Matrix Method (TMM) to model nonlinear optical phenomena in exciton-polariton microcavities. By incorporating the medium inhomogeneity and nonlinear response into the TMM framework, we developed a computationally efficient approach for simulating light propagation and interactions in nonlinear media. As an example, we focused primarily on exciton-polariton quantum well (QW) microcavities.

The theoretical derivation reformulates Maxwell's equations for structured media with nonlinear permittivity. Introducing a convolution term to account for spatially varying refractive indices, we derived a compact transfer matrix that incorporates nonlinear effects into the TMM. This formulation simplifies the analysis and enables efficient computation of eigenmodes, which are essential for studying resonance phenomena in polariton-based systems.

Leveraging the Fast Fourier Transform (FFT), the method achieves a time complexity of \( \mathcal{O}(N \log N) \), representing a significant improvement over diagonalization methods, which scale as \( \mathcal{O}(N^3) \). This scalability, especially for systems with large transverse Fourier mode counts \( N \), makes the method suitable for modeling complex optical systems with many spatial degrees of freedom and intricate nonlinear properties. The results confirm the method's applicability for solving complex problems in nonlinear photonics with reduced computational cost.

Numerical simulations demonstrate the method's capability to capture nonlinear effects, including spontaneous symmetry breaking (SSB) and self-focusing. SSB is observed as asymmetric intensity distributions in transmitted and reflected beams caused by the nonlinear QW, while self-focusing effects show beam narrowing due to intensity-dependent refractive index changes. These results validate the extended TMM for studying nonlinear light-matter interactions in polariton microcavities.

The method integrates with existing Maxwell equation frameworks and provides a versatile tool for analyzing polariton-based photonic devices. The ability to efficiently simulate nonlinear interactions supports the design and optimization of low-power optical devices, photonic circuits, and quantum photonic technologies. Future work may extend application of TMM to more complex geometries and material responses, broadening its relevance to nonlinear photonics.

\section*{ACKNOWLEDGEMENT}
This work was supported
by the European Union EIC Pathfinder projects ‘Quantum Optical Networks based on Exciton-polaritons’ (Q-ONE, ID 101115575) and ‘Neuromorphic Polariton Accelerator’ (PolArt, ID 101130304).

The code used in the simulations is available at \url{https://doi.org/10.5281/zenodo.15424046}.

\bibliography{bibliography}

    \section*{Appendix A.~~~~Boundary Conditions for the Microcavity with a Nonlinear Quantum Well} \label{sec:app_a}

Our aim is to solve the boundary conditions for a microcavity consisting of left and right Bragg mirrors and a nonlinear exciton-polariton quantum well in the center. The conditions expressed in the left- and right-propagating electric field basis are
    \begin{align}
        C^{-1} R \begin{bmatrix}
            1 & 0 \\
            i \widetilde{\epsilon}_{\rm eff} \star & 1
        \end{bmatrix} L C \begin{bmatrix}
            \widetilde{E}^+ \\ \widetilde{E}^-
        \end{bmatrix} &= \begin{bmatrix}
            \widetilde{E}^+_{\rm out} \\ 0
        \end{bmatrix},
    \end{align}
where \( C \) is given by Eq.~\eqref{eq:base_change}, while \( L \) and \( R \) are matrices corresponding to Eq.~\eqref{eq:tttm}. Substituting these expressions, we obtain
    \begin{align}
        \begin{bmatrix}
            1 & 1 \\
            n\cos\theta & -n\cos\theta
        \end{bmatrix}^{-1} \begin{bmatrix}
            R_1 & R_2 \\
            R_3 & R_4
        \end{bmatrix} \begin{bmatrix}
            1 & 0 \\
            i \widetilde{\epsilon}_{\rm eff} \star & 1
        \end{bmatrix} \nonumber \\ 
        \begin{bmatrix}
            L_1 & L_2 \\
            L_3 & L_4
        \end{bmatrix} \begin{bmatrix}
            1 & 1 \\
            n\cos\theta & -n\cos\theta
        \end{bmatrix} \begin{bmatrix}
            \widetilde{E}^+ \\ \widetilde{E}^-
        \end{bmatrix} &= \begin{bmatrix}
            \widetilde{E}^+_{\rm out} \\ 0
        \end{bmatrix}.
    \end{align}
Multiplication of the input fields by the change of basis and transfer matrices through the left half of the microcavity yields
\begin{align}
        &\begin{bmatrix}
            v_1 \\
            v_2
        \end{bmatrix} \equiv\begin{bmatrix}
            L_1 & L_2 \\
            L_3 & L_4
        \end{bmatrix} \begin{bmatrix}
            1 & 1 \\
            n\cos\theta & -n\cos\theta
        \end{bmatrix} \begin{bmatrix}
            \widetilde{E}^+ \\ \widetilde{E}^-
        \end{bmatrix} \nonumber\\
        &=\begin{bmatrix}
            (L_1 + n \cos\theta L_2) \widetilde{E}^+ + (L_1 - n \cos\theta L_2) \widetilde{E}^- \\
            (L_3 + n \cos\theta L_4) \widetilde{E}^+ + (L_3 - n \cos\theta L_4) \widetilde{E}^-
        \end{bmatrix}.
    \end{align}
Multiplication of the result by the nonlinear quantum well transfer matrix gives 
\begin{align}
    \begin{bmatrix}
        1 & 0\\
        i \widetilde{\epsilon}_{\rm eff} \star & 1
    \end{bmatrix}
    \begin{bmatrix}
        v_1 \\v_2
    \end{bmatrix} = 
    \begin{bmatrix}
        v_1\\
        i \widetilde{\epsilon}_{\rm eff} \star v_1 + v_2
    \end{bmatrix}.
\end{align}
Finally, multiplication by the transfer matrix through the right half of the microcavity and the inverse change of basis matrix brings us to
    \begin{align}
        &\frac{1}{2n \cos\theta}
        \begin{bmatrix}
            n \cos\theta & 1 \\
            n \cos\theta & -1
        \end{bmatrix} \begin{bmatrix}
            R_1 & R_2 \\
            R_3 & R_4
        \end{bmatrix} \begin{bmatrix}
            v_1\\
            i \widetilde{\epsilon}_{\rm eff} \star v_1 + v_2
        \end{bmatrix} \nonumber \\
        &=\frac{1}{2} \begin{bmatrix}
            \left(R_1 + \frac{R_3}{n\cos\theta}\right) v_1 + \left(R_2 + \frac{R_4}{n\cos\theta}\right) \left(i \widetilde{\epsilon}_{\rm eff} \star v_1 + v_2\right) \\
            \left(R_1 - \frac{R_3}{n\cos\theta}\right) v_1 + \left(R_2 - \frac{R_4}{n\cos\theta}\right) \left(i \widetilde{\epsilon}_{\rm eff} \star v_1 + v_2\right)
        \end{bmatrix}.
    \end{align}
    
We can now explicitly find the boundary condition for the reflected and transmitted fields. The incoming left-propagating field $\widetilde{E}^-$ satisfies the equation

\begin{align}
    &\left[\frac{n\cos\theta R_1 - R_3}{n\cos\theta R_2  - R_4} (L_1 + n \cos\theta L_2) +  (L_3 + n \cos\theta L_4) \right] \widetilde{E}^+ \nonumber \\
    &+\left[\frac{n\cos\theta R_1 - R_3}{n\cos\theta R_2  - R_4} (L_1 - n \cos\theta L_2) + (L_1 - n \cos\theta L_2) \right] \widetilde{E}^- \nonumber \\
    &+ i \widetilde{\epsilon}_{\rm eff} \star \left[ (L_1 + n \cos\theta L_2) \widetilde{E}^+ + (L_1 - n \cos\theta L_2) \widetilde{E}^-\right] = 0,
\end{align}
while the outgoing field reads
\begin{align}
   \widetilde{E}^+_{\rm out} =&  \Bigg[\left(R_1 + \frac{R_3}{n\cos\theta}\right)(L_1 + n \cos\theta L_2) \nonumber \\
   &\quad + \left(R_2 + \frac{R_4}{n\cos\theta}\right) (L_3 + n \cos\theta L_4)\Bigg]\widetilde{E}^+ \nonumber \\
   &+ \left(R_2 + \frac{R_4}{n\cos\theta}\right) i \widetilde{\epsilon}_{\rm eff} \star \left[(L_1 + n \cos\theta L_2) \widetilde{E}^+ \right] \nonumber \\
   &+\Bigg[\left(R_1 + \frac{R_3}{n\cos\theta}\right)  (L_1 - n \cos\theta L_2) \nonumber \\
   & \quad +  \left(R_2 + \frac{R_4}{n\cos\theta}\right)(L_3 - n \cos\theta L_4) \Bigg]\widetilde{E}^- \nonumber \\
   &+ \left(R_2 + \frac{R_4}{n\cos\theta}\right) i \widetilde{\epsilon}_{\rm eff} \star \left[(L_1 - n \cos\theta L_2) \widetilde{E}^-\right]
\end{align}

    \section*{Appendix B.~~~~Numerical and Theoretical Validation} \label{sec:app_b}
    To ensure the accuracy and reliability of the TMM simulations, we performed a series of validations against theoretical predictions and established simulation tools. Specifically, we compared the results from TMM with PLaSK for linear optical structures and utilized multiple theoretical approaches to verify various aspects of our implementation.
    
    Energy conservation was validated by calculating the Poynting vector at the input and output boundaries of the simulation domain. The Poynting vector \(\vec{S}\) \cite{born2013principles}, representing the energy flux of the electromagnetic field, is defined as
    \begin{align}
        \vec{S} = \frac{1}{2} \operatorname{Re}(\vec{E} \times \vec{H}^*),
    \end{align}
    where \(\vec{E}\) is the electric field, \(\vec{H}\) is the magnetic field, and the asterisk denotes the complex conjugate. By integrating the Poynting vector over the input and output surfaces, we ensured that the total energy entering the system matched the total energy exiting it.
    
\begin{figure}[H]
        \centering
        \includegraphics[width=0.47\textwidth]{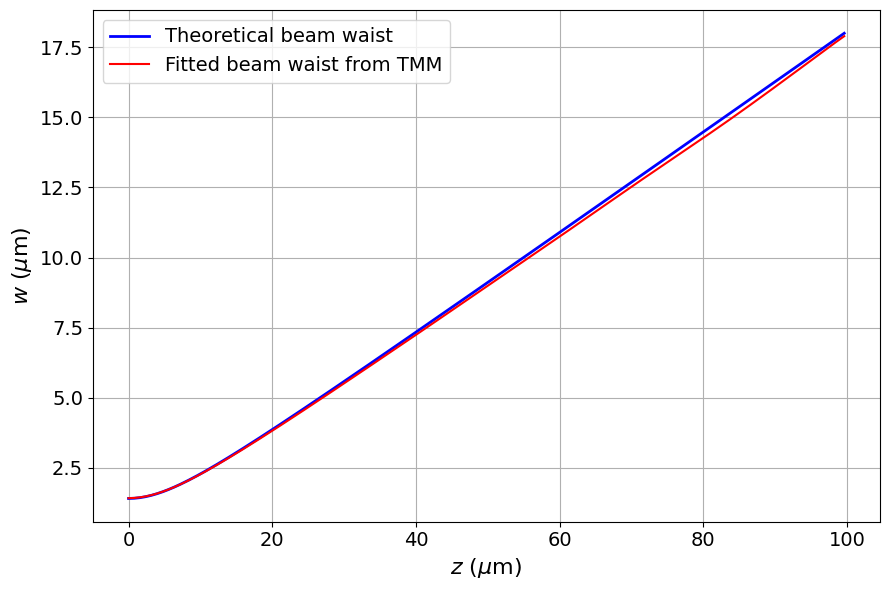}
        \caption{\justifying \fontsize{9pt}{11pt}\selectfont Comparison between the theoretical beam width evolution and fitted data obtained from TMM simulations.}
        \label{fig:app_ba}
    \end{figure}

    The propagation of Gaussian beams was verified by comparing the simulated beam waist evolution with the theoretical model. The width of a Gaussian beam, \(w(z)\), as a function of the propagation distance \(z\), is given by \cite{born2013principles}
    \begin{align}
        w(z) = w_0 \sqrt{1 + \left(\frac{z - z_0}{z_\text{R}}\right)^2},
    \end{align}
    where \(w_0\) is the beam waist (minimum beam width), \(z_0\) is the location of the beam waist, \(z_\text{R} = \frac{\pi w_0^2}{\lambda}\) is the Rayleigh range, and \(\lambda\) is the wavelength of the light. Simulated beam widths were fitted to this theoretical expression, and the results showed excellent agreement, as illustrated in Figure~\ref{fig:app_ba}.

\begin{figure}[H]
        \centering
        \includegraphics[width=0.49\textwidth]{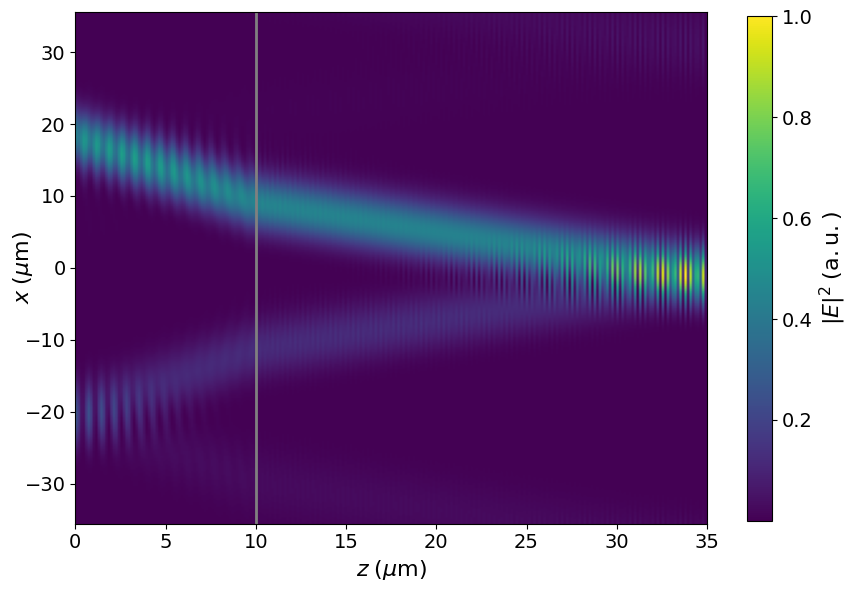}
        \caption{\justifying \fontsize{9pt}{11pt}\selectfont Simulation of Brewster angle incidence. The grey line represents the boundary between air ($n_1=1,d_1=10\, \mu m$) on the left and a semiconductor ($n_2=1.3,d_2=25\, \mu m$) on the right. The p-polarized light (Gaussian beam width $w_0=5\sqrt{2} \,\mu \text{m}$) exhibits zero reflected intensity at the Brewster angle $\theta_\text{B}\approx52.4^\circ$.} \label{fig:app_bb}
    \end{figure}

    Polarization effects were validated through a simulation of light reflection at the interface between two materials. The Brewster angle, \(\theta_\text{B}\), is the angle of incidence at which the reflected intensity of p-polarized light vanishes. For light incident from vacuum (\(n_1 = 1\)) onto a medium with refractive index \(n_2 = 1.3\), the Brewster angle \cite{brewster1815ix} is given by
    \begin{align}
        \tan \theta_\text{B} = \frac{n_2}{n_1}.
    \end{align}
    In our simulation, we verified that for p-polarized light, the reflected intensity dropped to zero at the Brewster angle. This result is consistent with theoretical predictions and demonstrates the ability of TMM to accurately simulate polarization-dependent phenomena. Figure~\ref{fig:app_bb} illustrates the field distribution at the Brewster angle, with the grey line marking the interface between the two materials.
    \vspace{0.1cm}
   
    These theoretical and numerical validations confirmed the accuracy of the TMM implementation in the linear regime, demonstrating its effectiveness in modeling complex photonic structures.

\end{document}